\documentstyle[epsfig,psfig]{mn}

\def\simgt{\ga}
\def\simlt{\la}

\begin{document}

\title[Detector Response Time \& Bolometric CMB Experiments]{The Effect
  of the Detector Response Time on Bolometric Cosmic Microwave
  Background Anisotropy Experiments}

\author[S.~Hanany, A.~H.~Jaffe and E.~Scannapieco]
{S.~Hanany,$^{1}$\thanks{\tt hanany@physics.berkeley.edu}
  A.~H.~Jaffe,$^{1}$\thanks{\tt jaffe@physics.berkeley.edu} and 
  E.~Scannapieco$^{1,2}$\thanks{\tt evan@physics.berkeley.edu}\\
  $^{1}$ The Center for Particle Astrophysics, University of California,
  Berkeley, 301 Le Conte Hall, Berkeley, CA, 94720-7304.\\
  $^{2}$ Physics Department, University of California, Berkeley, CA
  94720}

\maketitle

\begin{abstract}
  We analyze the effects of the detector response time on bolometric
  measurements of the anisotropy of the cosmic microwave background
  (CMB).  We quantify the effect in terms of a single dimensionless
  parameter $L$ defined as the ratio between the time the beam sweeps
  its own size and the bolometer response time.  As $L$ decreases
  below $\sim 2.5$ the point source response of the experiment becomes
  elongated. We introduce a window function matrix based on the
  timestream data to assess the effects of the elongated beam.  We
  find that the values of the window function matrix elements decrease
  slowly as a function $L$.  Our analysis and results apply to other
  cases of beam asymmetry. For the High Frequency Instrument on board
  the Planck Surveyor satellite we show that for a broad range of $L$
  the ability of the experiment to extract the cosmological parameters
  is not degraded. Our analysis enhances the flexibility in tuning the
  design parameters of CMB anisotropy experiments.
\end{abstract}


\section{Introduction}

The anisotropy of the cosmic microwave background (CMB) radiation
encodes a vast amount of information about structure formation in the
universe and about the values of the cosmological parameters 
(White, Scott \& Silk 1994). %\cite{white review}.
A number of groups are now making efforts to map the CMB at resolutions
of $10$--$20$~arcmin. Two satellite missions, NASA's MAP and ESA's
Planck Surveyor, are scheduled to be launched within the next decade.

In a substantial fraction of ongoing and planned experiments the
detector elements are bolometers. Bolometers are thermal detectors and
thus have a response time that depends on a variety of construction and
operation parameters \cite{richards review}.  Time constants of
bolometers currently employed on CMB experiments are on the order of 10
msec. The baseline design for Planck's High Frequency Instrument (HFI)
bolometers is between 2 and 5 msec.

The bolometer response time puts a practical constraint on the speed
with which a bolometric experiment can scan the sky.  Scanning  
too fast decreases the sensitivity of the experiment to structures
near the beam resolution. On the other hand, the scan speed and the
detector's $1/f$ noise knee determine the number of time-ordered
pixels over which the noise is uncorrelated. A faster scan speed
allows a more robust characterization of the instrumental noise and
better recovery of the sky signal over a larger angular range
\cite{tegmark1,tegmark2}.
Thus there are important trade-offs 
between the detector noise level, time constant, beam size and scan
speed that must be considered when optimizing the experiment.  In
the case of the Planck Surveyor instruments, for example, a balance
has to be found between the requirements of the low frequency
HEMT-based instrument (LFI) and the bolometric high frequency
instrument. The LFI prefers a fast scan speed because both its
detector response speed and $1/f$ noise knee are higher than for the
HFI. However, if the scan speed is too fast HFI's performance at high
resolution will degrade.

The window function matrix, and its trace, the window function, quantify
the sensitivity of a CMB anisotropy experiment as a function of angular
scale.  The full matrix contains information about all possible pixel
pairs; the simpler window function, $W_{\ell}$, reduces this to the RMS
anisotropy probed by the experiment.  A number of authors have analyzed
how the beam size and scan strategy determine the window function (see
White et al.\ \shortcite{white review} and references therein).  It has
traditionally been assumed that the bolometer response time can be
neglected. With a window function at hand, algorithms have been
developed to asses the accuracy with which a given experiment can
extract the cosmological parameters (Jungman et al.\ 1995; Bond,
Efstathiou \& Tegmark 1997; Zaldarriaga, Spergel \& Seljak 1997). 
%\cite{param est,zald,bond}.  
All of these calculations assume that the experiment illuminates the sky
with a symmetric beam and that the noise in the experiment is
uncorrelated, and hence that off-diagonal elements of the window
function matrix are unimportant.

In this paper we analyze how the window function matrix of a
bolometric experiment depends on the combination of bolometer response
time, beam size, and scan speed.  
When the detector response time cannot be neglected, the effective
shape of the beam becomes elongated and the form of the window
function matrix becomes a complicated function of the scan
strategy and pixelization. Furthermore, the standard techniques used to
assess the accuracy with which a given experiment can extract
cosmological parameters from the data are no longer strictly valid
since they assume that the beam is symmetric. We overcome the
difficulties associated with constructing a window function matrix
and assessing the performance of the experiment in the presence of an
asymmetric beam by introducing a window function constructed from data
points pixelized in the time domain.  We use this `temporal' window
function matrix to evaluate the window function as a function of the
detector response time and to estimate the magnitude of the necessary
corrections to the off-diagonal elements.  The discussion focuses on
the potential reduction in the window function matrix response near
the nominal resolution of the experiment.  Our analysis and results
will be applicable to any asymmetric beam elongation that decreases
the nominal resolution in one direction. As a special application, we
calculate the ability of Planck-HFI to extract cosmological
information as a function of different values of the instrumental
parameters.  

In Section \ref{section bolometer time constant} we derive the effect
of the detector response time on the point source response of the
experiment. We then derive the necessary changes to the zero lag
window function (Sections \ref{section temporal window function},
\ref{zero lag window function}) and assess
Planck-HFI's performance (Section \ref{section single point
cosmological parameters}). In Section \ref{two-point correlation
measurements} we concentrate on the full window function matrix in the
presence of the beam asymmetry introduced by the detector response
time. We discuss and summarize the results in Section \ref{section
discussion}.

\section{Bolometer Time Constant and Point Source Response}
\label{section bolometer time constant}

A bolometer is a thermal detector in which absorbed radiation is
converted to heat, causing a temperature change that is proportional 
to the absorbed energy. A thermistor is used to measure the temperature
change.  The temporal response of bolometers is an exponential; if the
bolometer temperature is $T_{0}$ at time $t<0$, then upon a step
function increase in input power at time $t=0$,
\begin{equation}
T(t>0) = T_{0} + \Delta T (1-\exp(-t/\tau)).
\end{equation} 
The time constant $\tau$ depends on the heat capacity of the absorbing
medium, on the detector's thermal conductivity to its mounting
structure, and on properties of the thermistor \cite{richards review}.
The exponential response time $\tau$ is a single-pole low-pass filter in
the frequency response of a bolometric experiment.  The filter is given
by
\begin{equation}
\label{equation single-pole time}
 {F}(\omega) = {1- i\, \omega \tau \over 
1 + (\omega\tau)^2} .
\end{equation}
The amplitude response of the filter has a $-3$ dB point\footnote{dB
= $20\log_{10}( \left| {F}\right|) $} 
at a frequency $f_{\rm bolo} = 1/(2\pi\tau)$.

In an observing strategy where the optical beam is scanned across the
sky the bolometer time constant can be viewed as a single-pole low
pass filter in the spatial domain. Since all of the effects we will be
discussing are relevant only at small angular scales, we approximate
the region of interest on the sky as flat and use Fourier transforms
on the plane.
If we write $\omega={\bf k \cdot v}$, where ${\bf k}$ is
the spatial wavenumber being probed, and ${\bf v}$ is the angular
velocity of the pointing on the sky, we get for the response in
${\bf k}$-space 
\begin{equation}
\label{equation single-pole space}
 {F}(k) = {1- i\, k v \tau\cos \theta_{kv} \over 
1 + (k v \tau\cos \theta_{kv} )^2} 
\end{equation}
where $\theta_{kv}$ is the angle between the vectors ${\bf k}$ and
${\bf v}$. $ {F}$ is a complex filter. Its amplitude response is
\begin{equation}
\label{amplitude}
\left| {F}(k)\right| = 
\frac{1}{\sqrt{1 + (kv\tau)^{2}\cos^{2}\theta_{kv}}},
\end{equation}
and its phase is 
\begin{equation}
\label{phase}
\phi = \tan^{-1}(- k v \tau\cos \theta_{kv}).
\end{equation}
The amplitude response of the filter has a $-3$ dB point at $k_{3dB} =
1/(v \tau \cos\theta_{kv})$ which depends on the instantaneous
scan speed $v$. From now on we will implicitly assume that, where
relevant, the scan speed is constant.  The approximation of a flat sky
is valid for regions smaller than $\sim 20 \times 20$ deg$^{2}$.
Within this approximation $k \simeq \ell$, where $\ell$ is the
multipole number of the $Y_{\ell m}$ spherical harmonics.

Assume that the optical system of a bolometric experiment produces a
symmetric beam $B({\bf r}, \sigma)$, where $\sigma$ is a measure of
the size of the beam (e.g., the width of a Gaussian beam), and ${\bf
r}$ locates it. The position ${\bf r}$ is defined relative to
coordinates located on the region of interest on the sky. If the beam
is swept across the sky at a constant velocity ${\bf v}$ the point
source response is the convolution of the beam and the temporal
response of the detector: \begin{equation}
\label{spatial space}
S({\bf r},{\bf v}, \tau,\sigma) = 
\int B(\left|{\bf r}-{\bf r}'\right|,\sigma)\, 
F({\bf r}',{\bf v},\tau)\, d^{2}{\bf r}';
\end{equation}
equivalently, 
\begin{equation}
\label{fourier space}
S({\bf r},{\bf v}, \tau,\sigma) = \int \exp (-i{\bf k}\cdot{\bf r})\, 
{  B}({\bf k},\sigma)\, {  F}({\bf k},{\bf v},\tau)
\frac{d^{2}{\bf k}}{(2\pi)^{2}}, 
\end{equation}
where, to simplify notation, we use  
$B({\bf k})$ and $F({\bf k})$ as the Fourier transforms of $B({\bf r})$
and $F({\bf r})$.  Since $F(k)$ 
is complex, the resulting response $S$ is both
attenuated and phase shifted compared to the case $\tau=0$.  A
dimensionless figure of merit that quantifies the level of
attenuation and phase shift is $L$, defined as
\begin{equation}
\label{define L}
L \equiv {  \sigma \over  v \tau } = \frac{\sigma / v }{\tau} =
\frac{f_{\rm bolo}}{f_{\rm scan}}. 
\end{equation}
Since $\sigma/v$ is the time it takes the beam to cross a distance
equal to its width, $L \gg 1$ corresponds to the case where the time
constant is short compared to this crossing time and spatial filtering
is negligible.  $L \ll 1$ corresponds to the case where the time
constant is much slower than the crossing time and spatial filtering
is expected to be significant. In the frequency (1/time) domain, $L$
is the ratio between the bolometer time constant low-pass filter
$f_{\rm bolo}$ and the width of the point source response with $\tau =
0$, $f_{\rm scan} \equiv v/(2\pi \sigma) $ \cite{hristov}.  For a
Gaussian beam $f_{\rm scan}$ is at the $1/e$ point of the point source
response.

\begin{figure}
\centerline{\psfig{file=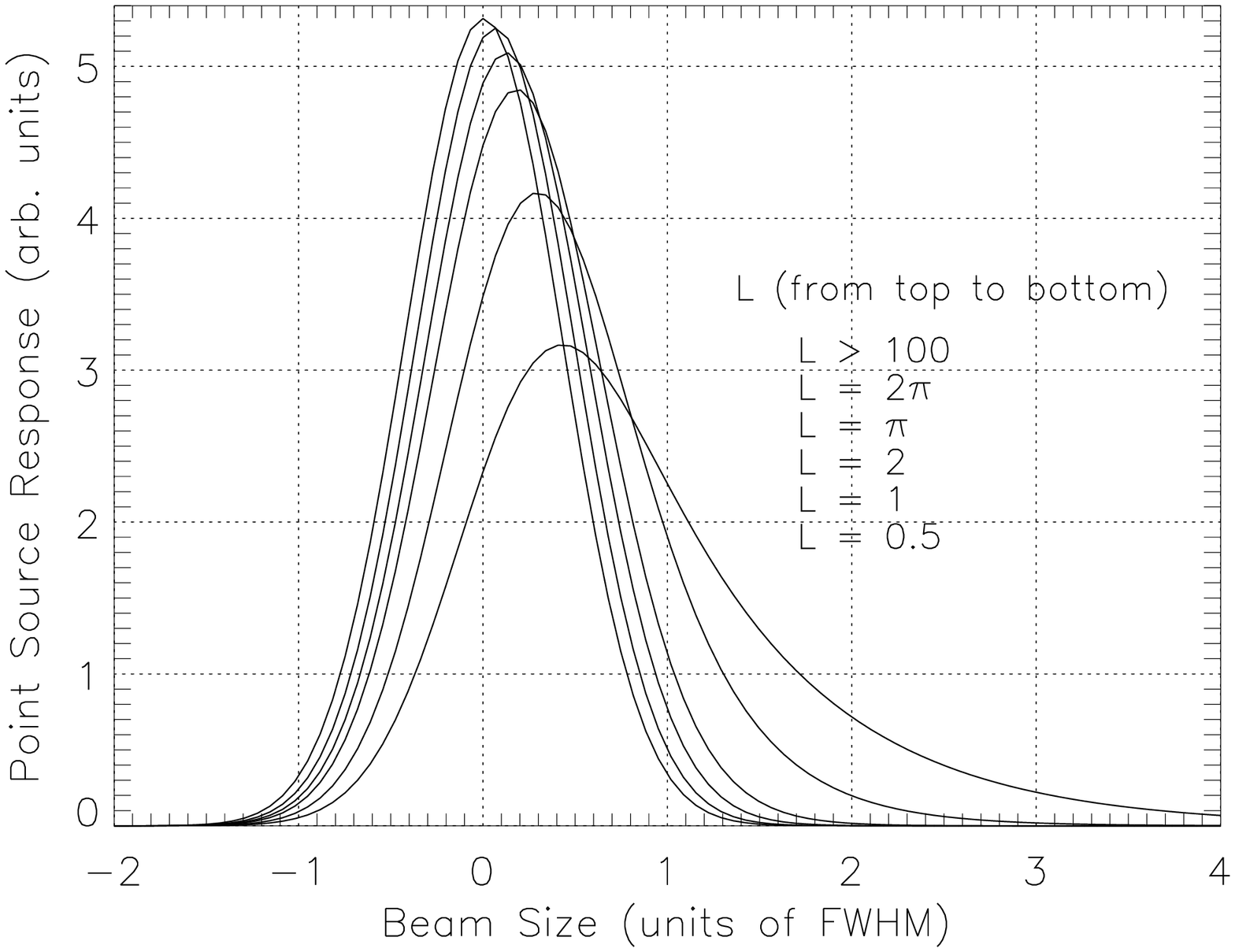,width=3.0in,angle=90}}
\centerline{\psfig{file=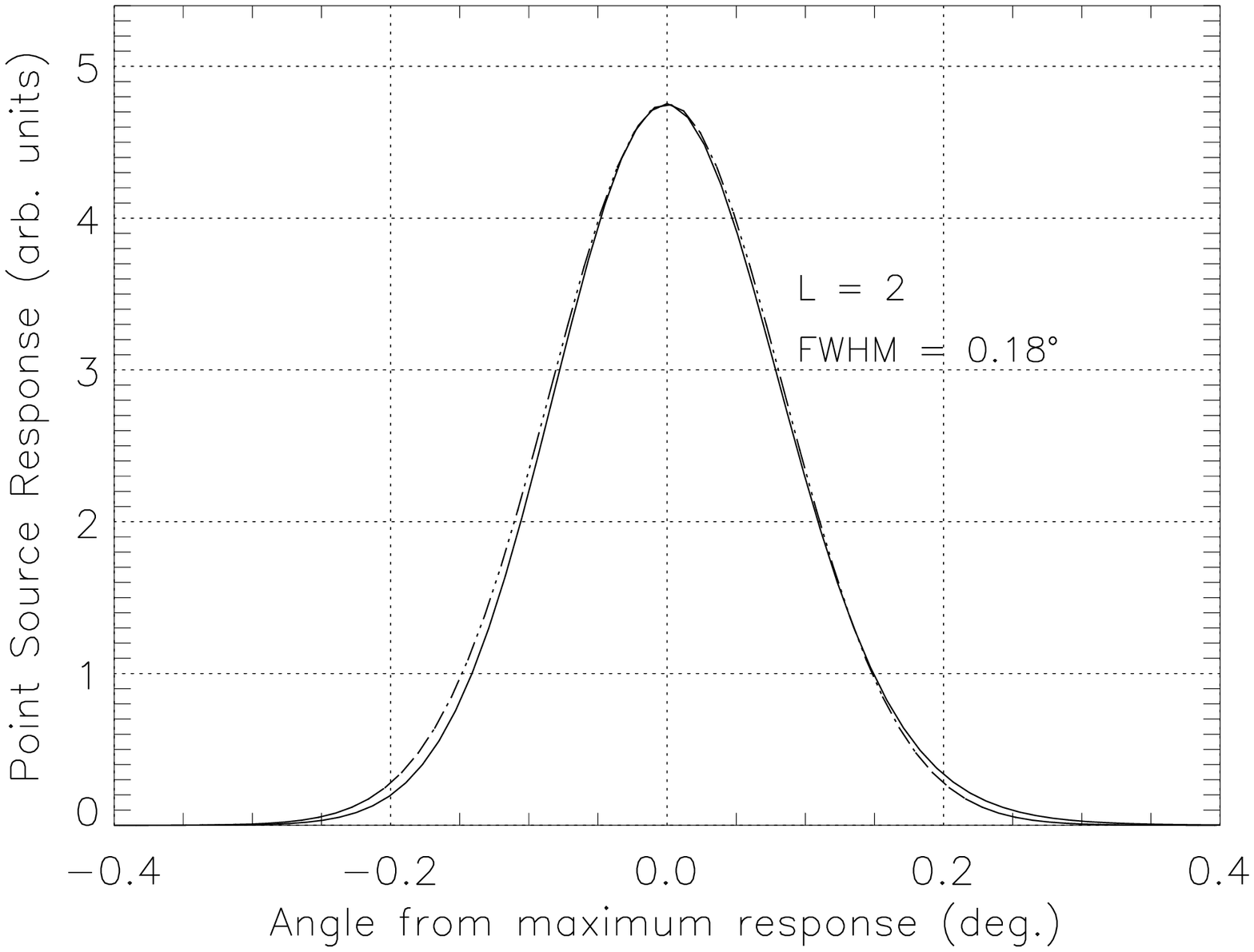,width=3.0in,angle=90}}
\caption{Top panel: 
  The point source response of a Gaussian beam as a function of $L =
  \sigma/ (v \tau)$, where $\sigma$ is the width the beam, $v$ is the
  scan speed, and $\tau$ is the bolometer time constant.  For $L \gg
  1$ there is little change in the shape of the beam.  For $L \leq 2$
  the point source response is phase shifted and attenuated.  Bottom
  panel: For values of $L > 2$ the effective shape of the beam in the
  direction of the scan is well approximated by a Gaussian with a
  width $\sigma_{1d}^{2}\simeq\sigma^2+(v\tau)^2$.  Shown are the
  effective shape (full curve) and the Gaussian fit (dash-dot) for $L
  =2$.}
\label{figure one}
\end{figure}

The upper panel in Figure \ref{figure one} shows the point source
response of an initially Gaussian beam for a range of $L$ values. For
a nominal two-dimensional Gaussian beam,
\begin{equation}
B({\bf r},\sigma) = 
\frac{1}{2 \pi \sigma^{2} }  
e^{-{r^{2}}/({2\sigma^{2}})} 
\end{equation}
and 
\begin{equation}
{  B}({\bf k},\sigma) =
e^{-{k^{2}\sigma^{2}}/{2} } .
\end{equation}
As expected, small values of $L$ result in a larger effective beam in
the direction of the scan.  The shape in a direction perpendicular to
the scan direction is not affected. Overall, the larger effective beam
size will result in a degraded sensitivity to spatial structures which
are on the scale of the nominal resolution.

For values of $L \simgt 2$ the spatial filter (Eq.\ \ref{amplitude}) can
be approximated as a Gaussian. When combined with a nominal 
Gaussian beam of width $\sigma$ the Fourier transform, Eq.
\ref{fourier space}, can be solved exactly, and one finds that 
the effective beam in the direction of the scan has acquired 
an additional 
Gaussian of width $\sigma_{\rm bolo}^2 \simeq (v\tau)^2$.
The total beam width in the direction of the scan is now
\begin{equation}
\label{new 1d width}
\sigma_{1d}^{2} \simeq \sigma^{2} +(v\tau)^{2} = \sigma^{2}(1+L^{-2}).
\end{equation}
The lower panel of Fig.\ \ref{figure one} shows the effective beam
shape and the Gaussian that approximates it for $L=2$.  The resulting
2-dimensional beam is elliptical with a width $\sigma_{1d}$ in the
direction of the scan and $\sigma$ in a direction perpendicular to the
scan.

\section{Window Function and Cosmological Parameter Estimation}
\label{window function and cosmological parameters}

We now quantify the effect of the enlarged asymmetric beam on the
window function of the experiment and its effect on the estimation of
cosmological parameters. We ignore the spatial phase shifts of the
signal. In most real situations the phase shift will either be small
compared to the beam size or the need to accurately reconstruct the
beam location (for e.g., making an accurate map) will force
experimenters to determine and correct for it.

\subsection{Temporal Window Function}
\label{section temporal window function}

We begin with the temporal data stream which consists of data
labeled by both position on the sky and time.  The signal at a given
time $t=1\ldots N_t$ is a convolution of the beam pattern with the
underlying sky signal:
\begin{equation}
  s_t = \int d^2{\bf\hat x} \; {\cal F}_t({\bf\hat x})
  {\Delta T\over T}({\bf\hat x}) = \sum_{\ell m} {\cal F}_{t\ell m}
  a_{\ell m}
\end{equation}
where we have transformed to spherical harmonics in the second
equality. The kernel ${\cal F}$ encodes the beam shape and its location on
the sky at time $t$.  If we consider small areas of sky, on which the
time-constant effects will be relevant, we can again consider the sky to be
flat and approximate the spherical harmonics with Fourier
transforms. Then the signal is
\begin{equation}
s_t = \int {d^2{\bf k}\over(2\pi)^2} a({\bf k}) {\cal F}_{t}({\bf k}),
\end{equation}
with 
\begin{equation}
\langle a({\bf k})a({\bf k'})\rangle = (2\pi)^2 \delta^2({\bf k}+{\bf k'})C(k),
\end{equation}
and 
\begin{equation}
C_\ell \simeq \left.C(k)\right|_{\ell \simeq k}.
\end{equation}

The temporal signal correlation matrix is
\begin{equation}
S_{tt'} =\langle s_t s_{t'}\rangle =
\int d\ln k\; W_{tt'}(k) {k^2 C(k)\over2\pi}
\end{equation}
with the temporal window function matrix
\begin{equation}
  W_{tt'}(k)=\int\;{d\theta_k\over2\pi} e^{-i k r_{tt'}\cos\theta_k}
  U_{tt'}({\bf k}) B^2(k)
\end{equation}
where we have split off the effect of a symmetric beam $B(k)$ and
other asymmetric parts $U$. Here, $r_{tt'}$ is the angular distance
between pixels observed at times $t$, $t'$. An experiment with $U =
1$, (i.e., a symmetric beam, and no chopping), has
\begin{equation}\label{Wzero}
  W_{tt'}(k)=J_0(kr_{tt'})B^2(k).
\end{equation}
For $t=t'$ we get the zero-lag window function
\begin{equation}\label{Wzerozero}
  W_{tt}(k) = B^2(k).
\end{equation}
With the spatial asymmetry of the beam, induced by the bolometer time
constant (Eq.\ \ref{amplitude}), the temporal
window function matrix becomes
\begin{eqnarray}\label{Wtau}
  W_{tt'}(k)&=& B^2(k)\int\;{d\theta_k\over2\pi} e^{-i k r_{tt'}\cos\theta_k}
  \nonumber\\
  &&\times\left[1+(k v \tau)^2 \cos^2(\theta_k-\theta_t)\right]^{-1/2}
  \nonumber\\
  &&\times\left[1+(k v \tau)^2 \cos^2(\theta_k-\theta_{t'})\right]^{-1/2}, 
\end{eqnarray}
where $\theta_t$, $\theta_{t'}$ are the angles of the velocity vector
$\bf{v}$ at times $t$ and $t'$.

\subsection{Zero Lag Window Function}
\label{zero lag window function}

\begin{figure}
\centerline{\psfig{file=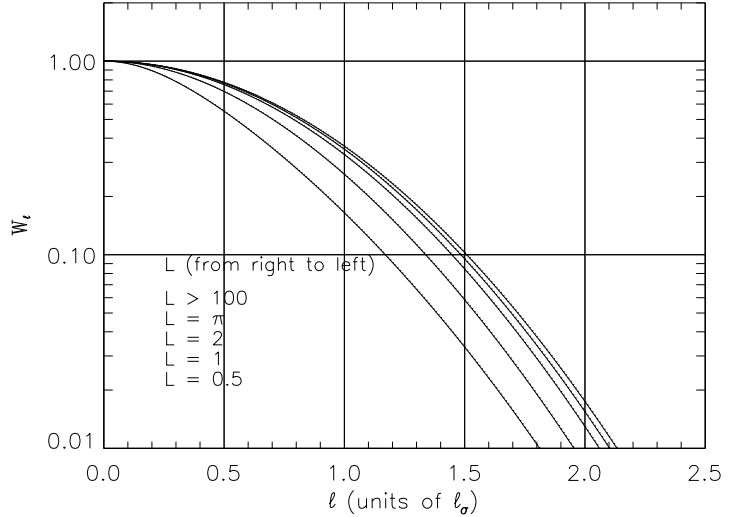,width=3.0in,angle=90}}
\caption{The zero lag window function, Eq.\ \ref{single point window}, with an
  initially Gaussian beam for a range of $L$ values. The horizontal axis is 
  in units of $\ell_{\sigma} = 1/\sigma$. }
\label{figure filters2}
\end{figure}

To obtain the traditional window function of the experiment, which
encodes contributions only to the RMS signal, we set $t=t'$ in
Eq.~\ref{Wtau}, giving
\begin{eqnarray}
\label{single point window}
  W_{tt}(k) &=& B^2(k)\int\,{d\theta_k\over2\pi}\,
  \left[1+(k v \tau)^2 \cos^2(\theta_k-\theta_t)\right]^{-1}\nonumber\\
  &=&        B^2(k) {1\over\sqrt{1+(k/k_\tau)^2}}
\end{eqnarray}
where
\begin{eqnarray}
k_\tau \simeq \ell_\tau \equiv
\frac{1}{v \tau } = 
\frac{1.8\cdot10^{5}}{\pi}
\left(\frac{1 {\rm ms}}{\tau}\right)
\left(\frac{1 {\rm deg/sec}}{v}\right).
\end{eqnarray}
The original window function, which was composed of the beam spatial
filter, $W_{\tau=0}(k)=B^2(k)$, is now multiplied by an additional
single-pole low-pass filter,
\begin{eqnarray}
  \label{equation define r}
  R(\ell) &\simeq& R(k)=W(k)/W_{\tau=0}(k)\nonumber\\
  &=&[1+(k/k_\tau)^2]^{-1/2} \simeq [1+(\ell/\ell_\tau)^2]^{-1/2}.
\end{eqnarray}
The $-3$ dB point of $R(\ell)$ is at $\ell = \ell_{\tau}$.  The beam
filter has a low-pass cut-off at $\ell_{\sigma} \equiv 1/\sigma$. The
shape of the filter depends on the shape of the beam. ( A Gaussian
beam with width $\sigma$ has an exponential low pass $
B^{2}(k) = \exp{(-\ell^{2}\sigma^{2})}$.)
Thus the effects of beam elongation will be important when 
\begin{equation}
\ell_{\tau} \leq \ell_{\sigma} \Rightarrow 
L = \frac{\ell_{\tau}}{\ell_{\sigma}} \simlt 1.
\end{equation}
Figure \ref{figure filters2} shows the zero lag window function for an
originally Gaussian symmetric beam and for various values of $L$.

We point out that the window function calculated from the timestream,
Eq.\ \ref{single point window}, is the {\em same} as the one
calculated from a map,
\begin{equation}
W_{\ell} = { 1 \over N_{p}} \sum_{N_{p}} W_{pp}(\ell),
\end{equation}
where $p$ denotes a pixel location on the sky.  The window function,
in any representation, just measures the sensitivity to the RMS
temperature, and does not take into account pixel-pixel correlations.

For a reasonable range of $v$ and $\tau$, $\ell_\tau\simgt 800$, 
below which most of the cosmological information resides in CDM-like
models. 
% this will become clearer in the discussion, Section \ref{section
% discussion}.
% the reason for this choice of numerical value 
For these cases, and for $\ell \simlt \ell_{\tau}$, the single-pole
filter $R_{\ell}$ can be usefully approximated by a Gaussian.  By Taylor
expanding $(1 + \ell^2/\ell_\tau^2)^{-1/2}$ and matching with the first
term of a similar expansion for a Gaussian we find that
\begin{equation}
  \label{symmetric window}
  W_{g}(\ell) \simeq B^2(\ell) e^{- \ell^2 \sigma'^{2}} 
  = e^{- \ell^2 (\sigma^{2} + \sigma'^{2})} 
\end{equation}
where
\begin{equation}
  \sigma'^{2} \equiv \frac{1}{2 \ell_\tau^2}=\frac{(v\tau)^2}{2},
\end{equation}
and we have specialized to a Gaussian beam in the equality.  This can
be thought of as approximating the nearly ellipsoidal effective beam by
a circular Gaussian beam with a slightly larger width.  This new width,
\begin{equation}
\label{new 2d width}
\sigma_{2d}^{2} \simeq \sigma^{2} + \frac{(v\tau)^{2}}{2} =
\sigma^{2} ( 1+L^{-2}/2 )
\end{equation}
is larger than the original width $\sigma$ and smaller than the one
dimensional effective width that we found in the previous section; see
Eq.\ \ref{new 1d width}.  That is, we find a degradation in the
window function 
response compared to the original $\sigma^2$ width, but not as severe
as indicated by the beam elongation in the scan direction,
$\sigma^{2}_{1d}=\sigma^2(1+L^{-2})$.  This is because beam elongation
occurs primarily in one direction whereas the window function encodes
information from all spatial directions. Generally, as can also be
observed by Comparing Figs.\ \ref{figure one} and \ref{figure
filters2}, the effects of small values of $L$ are stronger in the
spatial domain than they are on the window function.

The effective Gaussian filter $\exp(-\ell^{2}(\sigma^{2}+\sigma'^{2}))$
is the geometric mean of two Gaussian filters corresponding to two
symmetric beams \cite{white},
\begin{equation}
B^{2}(k) = \sqrt{ B^{2}_{\sigma}(k) B^{2}_{\sigma_{1d}} (k)}.
\end{equation}
One has a width $\sigma_{1d}$ corresponding to the effective beam in
the scan direction and the other has a width $\sigma$ corresponding to
the nominal beam size. The Gaussian approximation is good to within $10\%$
for all $\ell$ less than $.88 \, \ell_\tau$. As an example,
$\ell_{\tau}= 1680$ for $\tau = 5$ msec.  and $v = 6$ deg./sec. which
are the time constant for Planck's 100 GHz bolometric channel, and the
satellite rotation rate, respectively, and thus the approximation is
valid for $\ell \leq 1480$.

\subsection{Estimation of Cosmological Parameters}
\label{section single point cosmological parameters}

Given the window function $W(k)$ (Eq.\ \ref{single point window}), we
use the apparatus that has been developed by Jungman et al.\ 
\shortcite{param est} and others \cite{bond,zald} to estimate its effect
on the determination of cosmological parameters. We specialize the
discussion to the experimental configuration of the High Frequency
Instrument (HFI) on board the Planck Surveyor satellite.

\begin{table}%[htbp]
 \begin{center}
   \leavevmode
\begin{tabular}{|l|lll|}
  \hline $h=0.5$ SCDM &       $L_{0}$    &$L_{0}/2$  &$L_{0}/5$\\\hline
${H_0}$       [rel]                  & 0.021  & 0.021  & 0.024 \cr    %% DONE
$\Omega_0$                           & 0.002  & 0.002  & 0.002 \cr
$\Omega_{\rm vac}$                   & 0.049  & 0.050  & 0.058 \cr
${\Omega_{b}h^2}$       [rel]    & 0.006  & 0.006  & 0.007 \cr
$\Omega_{\nu}$                       & 0.018  & 0.018  & 0.020 \cr
%BP [unused]                         & 0.014  & 0.014  & 0.016 \cr
$n$                                  & 0.006  & 0.006  & 0.007 \cr
tensor/scalar                        & 0.095  & 0.095  & 0.098 \cr
$\tau$                               & 0.155  & 0.155  & 0.158 \cr
  \hline\hline $\Omega=0.33$, $h=0.6$ open CDM &$L_{0}$ &$L_{0}$/2 &$L_{0}$/5
  \\\hline
${H_0}$       [rel]                  & 0.011  &  0.011 & 0.015\\
$\Omega_0$                           & 0.001  &  0.001 & 0.001\\
$\Omega_{\rm vac}$                   & 0.016  &  0.017 & 0.021\\
${\Omega_{b}h^2}$       [rel]    & 0.008  &  0.008 & 0.009\\
%Omega_mnu                           & 0.000  &  0.000 & 0.000\\
%CLBpowh                             & 0.015  &  0.015 & 0.017\\
$n$                                  & 0.003  &  0.003 & 0.003\\
%tensor/scalar                       & 0.000  &  0.000 & 0.000\\
$\tau$                               & 0.053  &  0.053 & 0.054\\
%H0_rel_kfix                         & 0.014  &  0.014 & 0.018\\
%H0_rel_vfix                         & 0.003  &  0.003 & 0.004\\
%omcdm_rel                           & 0.006  &  0.006 & 0.008\\
%Omcurv                              & 0.015  &  0.015 & 0.019\\
%omk_kfix                            & 0.005  &  0.005 & 0.007\\
%omv_vfix                            & 0.000  &  0.000 & 0.000\\
  \hline\hline $\Omega=0.33$, $\Omega_\Lambda=0.66$, $h=0.7 $ CDM & $L_{0}$ &
 $L_{0}/2$ &$L_{0}/5$ \\\hline
${H_0}$       [rel]                  & 0.011  & 0.011  & 0.013\\
$\Omega_0$                           & 0.002  & 0.002  & 0.002\\
$\Omega_{\rm vac}$                   & 0.030  & 0.030  & 0.035\\
${\Omega_{b}h^2}$       [rel]    & 0.007  & 0.007  & 0.008\\
%Omega_mnu                           & 0.007  & 0.007  & 0.008\\
%CLBpowh                             & 0.015  & 0.015  & 0.017\\
$n$                                  & 0.007  & 0.007  & 0.007\\
tensor/scalar                        & 0.078  & 0.078  & 0.080\\
$\tau$                               & 0.146  & 0.147  & 0.152\\
%H0_rel_kfix                         & 0.012  & 0.012  & 0.014\\
%H0_rel_vfix                         & 0.006  & 0.006  & 0.006\\
%omcdm_rel                           & 0.027  & 0.028  & 0.030\\
%                                                      
%omk_kfix                            & 0.002  & 0.002  & 0.002\\
%omv_vfix                            & 0.000  & 0.000  & 0.000\\
  \hline\hline
\end{tabular}
\caption{The three columns of numbers quantify the accuracy with 
  which the Planck Surveyor HFI will be able to determine a variety of
  cosmological parameters as a function of $L =\sigma/v\tau $, where
  $\sigma$ is the width of a Gaussian beam, $v$ is the scan speed, and
  $\tau$ is the detector time constant. The parameter $L_{0} = 2.67$ is
  appropriate for the baseline design of the HFI.  The calculation
  assumes 4 frequency channels at 100, 143, 217, 353 GHz, Gaussian beams
  with FWHM 10.6, 7.4, 4.9, 4.5 arcmin, nominal sensitivities of 4.9,
  4.8, 9.7, $40 \mu K$/pixel., respectively, and a scan rate of 6
  deg./sec. The notation `rel' is for fractional errors with respect
  to the model. Due to a degeneracy between $\Omega_{\rm vac}$ and
  $\Omega_0$, we hold the curvature, $1-\Omega_0-\Omega_{\rm
    vac}-\Omega_{b}-\Omega_\nu$, fixed for the error determination,
  except in the determination of $\Omega_0$ itself, for which we fix
  $\Omega_{\rm vac}$.}
\label{table cosmological parameters}
\end{center}
\end{table}

Each of the panels of Table \ref{table cosmological parameters}
represents a different cosmological model. For each of the models we
calculate the error in the determination of a variety of cosmological
parameters for three values of the parameter $L$.  $L_{0} = 2.67$ is
chosen to represent the value presently designed for the HFI
instrument.

Within the bounds of validity of the analysis (see below), we find
that for the case of the HFI choosing $L$ values which are a factor of
2 or 5 smaller than the nominal $L_{0}$ would not significantly affect
the ability to extract the values of the cosmological parameters. Note
that the effect of the smearing is almost completely negligible for
$L\simgt L_0/2\simeq 1.33$ and begins to be noticeable with the three
significant figures in the table at $L\simeq L_0/5\simeq0.5$.  We
defer a more detailed discussion of this result to Section
\ref{section discussion}.

The results in Table \ref{table cosmological parameters} should be read
with some caution.  The forecasts of parameter errors assume a form for
the covariance matrix (more precisely, the Fisher Matrix; see
Section~\ref{section off diagonal elements}) which is appropriate for an
experiment with a symmetric beam and uniform, uncorrelated pixel errors.
Since in our case the beam is manifestly not symmetric we expect these
forecasts to be only approximate. Furthermore, our analysis considers
quantities in the time domain.  The relationship between pixel- and
time-domain is not completely straightforward in the presence of beam
asymmetry; this is a topic of current research.  In the limit of many
crossings through the same pixel in many different directions, the
requirement for an effectively symmetric beam is satisfied, if we
consider quantities in the pixel, rather than time, domain. In other
cases it will not be satisfied and care must be taken in interpreting
the results in Table \ref{table cosmological parameters}.  White
\shortcite{white} has noted that it may be possible to get a
pixel-domain window function that is narrower than the time stream
window function that we use here by certain signal-to-noise weighting
techniques. More generally, it has been noted that the least-squares
method of Wright \shortcite{wright}, which is a variant of the method
originally used to make the COBE/DMR maps (see also
\cite{tegmark1,tegmark2}), produces a map containing all of the sky
information of the experiment. The method, as implemented in these
references, assumes a known timestream noise matrix $N_{tt'}$ and uses
it to estimate the temperature at a pixel $p$ given measurements of the
temperature at $p$ taken at different times. With an asymmetric beam,
individual measurements taken at $p$ no longer measure quite the same
quantity due to the differing beam orientations. This can be accounted
for at the cost of greatly complicating the matrices used in the
solution to the problem. Such a treatment is beyond the scope of this
paper. For the results of Table \ref{table
  cosmological parameters} we used timestream quantities and the RMS
window function $W(k)$.  We will return to the question of beam
asymmetry and assess its impact on the window function matrix, and on
the ability to extract cosmological parameters, in the following
section.

Finally,  the standard analysis, which leads to the results in  Table
\ref{table cosmological parameters}, makes several additional 
assumptions that need to be highlighted. The actual likelihood as a
function of the cosmological parameters will be highly non-Gaussian,
but the calculations assume Gaussianity (at least near the peak of the
distribution). In addition, the results assume that the removal of
foregrounds and systematic effects proceeds without appreciable
degradation of the satellite's sensitivity.

\subsection{Full Window Function Matrix}
\label{two-point correlation measurements}

The analysis in terms of the zero-lag window function is an
approximation that neglects pixel-pixel, or beam-beam correlations.
We now assess whether sweeping an effectively asymmetric beam across
the sky significantly affects this approximation. We first parameterize
the relative power spectrum errors $\delta C_{\ell}/ C_{\ell}$ in
terms of the symmetric beam errors and $a$. The parameter $a$
quantifies the `average' ratio in the beam-beam correlations between
the symmetric and asymmetric beam cases. We then assess the magnitude
of $a$.

\subsubsection{Off-Diagonal Elements and Parameter Estimation}
\label{section off diagonal elements}

The (ensemble average) error on the cosmological parameters $b_i$ is given by
\begin{equation}
  \langle \delta b_i \; \delta b_{i'} \rangle = F^{-1}_{ii'},
\end{equation}
where $F_{ii'}$ is the {\em Fisher Matrix} given by 
\begin{equation}
  \label{fishmat}
  F_{ii'} = {1\over2}{\rm Tr}\left( C^{-1} {\partial C\over\partial b_i}
 C^{-1} {\partial C\over\partial b_i'} \right).
\end{equation}
Here $C=C_{pp'}$ is the total correlation matrix (in pixel space,
although the arguments hold for the timestream data as well) of
the data in question, which has contributions from both signal and
noise: $C=C(S,N)=S+N$. 
In particular, the signal part of the matrix is given by
\begin{equation}
  S_{pp'} = \sum_\ell {2\ell+1\over 4\pi} W^\ell_{pp'} C_\ell.
\end{equation}
Thus, if we parameterize by the power spectrum itself the derivatives
are just 
\begin{equation}
    {\partial C\over\partial C_\ell}=
    {\partial S\over\partial C_\ell}={2\ell+1\over 4\pi} W_{pp'}^\ell.
\end{equation}
Here we write the full window function matrix as $W_{pp'}^\ell$
and the usual `window function' as $W_\ell=(1/N_p) {\rm
  Tr}(W^\ell)=(1/N_p)\sum_p W_{pp}^\ell$.
The Fisher matrix for any other set of parameters can be recovered from
that of the power spectrum by the appropriate Jacobian matrix:
\begin{equation}  
  F_{ii'}=\sum_{\ell\ell'}{\partial C_\ell\over\partial b_i} F_{\ell\ell'} 
  {\partial C_{\ell'}\over\partial b_{i'}} .
\end{equation}

Now we can determine the effect of a window function with an
asymmetric beam on parameter errors. We refer all the requisite
quantities to those with a symmetric beam. Define $W_{pp'}^\ell
\sim r_\ell{\overline W}_{pp'}^\ell$, where ${\overline W}_{pp'}^\ell$ is the
window function for a symmetric beam and $r_\ell\sim R(\ell)$ is an
appropriate pixel average of the ratio $R$ we defined in Eq.\
\ref{equation define r}. 
Also define
$S_{pp'}=a{\overline S}_{pp'}$ where again ${\overline S}$ is the
quantity for a symmetric beam, and $a$ is thus a weighted average of
the $r_\ell$. Since the form of $W$ will not be the same as that of
${\overline W}$, these ratios are of course only approximations.  With
these definitions we write the Fisher matrix in terms of
symmetric-beam quantities:
\begin{eqnarray}
  F_{\ell\ell'}&\sim&
  {1\over2}{\rm Tr}\left[
    (a {\overline S}+N)^{-1} r_\ell {\overline W}^\ell 
    (a {\overline S}+N)^{-1} r_{\ell'} {\overline W}^{\ell'}\right]  
\nonumber\\  &=& {r_\ell r_{\ell'}\over a^2}{1\over2}{\rm Tr}\left[ 
    ({\overline S}+N/a)^{-1} {\overline W}^\ell 
    ({\overline S}+N/a)^{-1} {\overline W}^{\ell'} \right]\nonumber\\
    &=& {r_\ell r_{\ell'}\over a^2} F_{\ell\ell'}({\overline S},N/a).
\end{eqnarray}
That is, we can write the Fisher matrix as approximately a factor times
the Fisher matrix for a symmetric beam, with the noise variance degraded
by $1/a$. We might expect this factor, $r_\ell r_{\ell'}/a^2$,
to be of order one.

In particular, for an experiment with uniform noise, we can write the
Fisher matrix for an experiment observing a fraction $f_{\rm sky}$ of
the sky with a symmetric beam $B_\ell$ as (approximately)
\begin{eqnarray}
  F_{\ell\ell'}\simeq (\ell+1/2)f_{\rm sky} \left[
    C_\ell + {1\over{\overline w}B^2_\ell}\right]^{-2}\delta_{\ell\ell'}.
\end{eqnarray}
Here, ${\overline w}={\rm Tr}(N^{-1})$ is the total weight of the
experiment, so the effect of the beam asymmetry is to make the replacement
${\overline w}\to a{\overline w}$; when the noise term dominates, the
error on $C_\ell$ is increased by $1/{a}$.

\subsubsection{Off-Diagonal Elements of the Window Function Matrix}
\label{off diagonal elements}

We now assess the magnitude of the off-diagonal elements of the window
function matrix in the presence of beam elongation due to small values
of the parameter $L$.  We will be interested in the ratio of
Eq.~\ref{Wtau}, which is the general expression for the window
function with an asymmetric beam, to Eq.~\ref{Wzero}, which assumes a
symmetric beam
\begin{eqnarray}
\label{eq:WtautoWzero}
R_{tt'}(k) & =  & W_{tt'}/\left\{B^2(k) J_0(kr_{tt'})\right\} \nonumber\\
           & =  & \left[1\over J_0(kr_{tt'})\right] 
\int\;{d\theta_k\over2\pi} e^{-i k r_{tt'}\cos\theta_k} \nonumber \\
 & & \times  \left[1+(k v \tau)^2 \cos^2(\theta_k-\theta_{t})\right]^{-1/2}
   \nonumber \\
 & & \times  \left[1+(k v \tau)^2 \cos^2(\theta_k-\theta_{t'})\right]^{-1/2}.
\end{eqnarray}
Note that the original symmetric beam $B(k)$ drops out of this
expression.  When this ratio is small the beam elongation is
having a large effect.  While this ratio cannot be calculated analytically,
we can gain much insight into the time-time correlations by considering 
the range in which $k v \tau = \ell/\ell_{\tau}$ is less than one. We then 
compute the integral numerically for larger values of $kv\tau$. 

Expanding Eq.\ \ref{Wtau} in terms of $\alpha \equiv (k v \tau)^2,$ 
redefining the zero point of the angular integral,
and working to order $\alpha^2$ we find
\begin{eqnarray}
\label{eq:angular average}
R_{tt'}(k) & = & 
{1 \over J_0(kr_{tt'})}
\int_0^{2 \pi} \frac{d \theta}{2 \pi} e^{i k r \sin \theta} 
\bigg\{ 1 \nonumber\\
& - & \frac{\alpha}{2} 
\left[ \sin^2(\theta_t - \theta) + \sin^2(\theta_{t'} - \theta) \right]
\nonumber \\
 & + & \alpha^2  
\big[ \frac{3}{8} \sin^4(\theta_t - \theta) \nonumber\\
&&\phantom{\alpha^2 \big[} 
+ \alpha^2 \frac{1}{4} \sin^2(\theta_t - \theta) \sin^2(\theta_{t'} -\theta)
\nonumber\\
   &&\phantom{\alpha^2 \big[}   + \frac{3}{8} \sin^4(\theta_{t'} - \theta) \big]
\bigg\}.
\end{eqnarray}
By contour integration, this can be broken up into a sum of spherical
Bessel functions:
\begin{eqnarray}
R_{tt'} J_0(kr) & = & J_0(kr) \nonumber\\
&-&\frac{\alpha}{2} \left[ J_0(kr) 
- \frac{\cos(2 \theta_t) + \cos(2 \theta_{t'})}{2} J_2(kr) \right]
\nonumber \\
& + & \frac{\alpha^2}{32}
\Bigg\{
\left( \frac{3}{2} \cos^4 \theta_t +  \cos^2 \theta_t \cos^2 \theta_{t'} 
+ \frac{3}{2} \cos^4 \theta_{t'} \right)\nonumber\\
&&\phantom{\frac{\alpha^2}{32}\Bigg\{}
\times\left[ 3 J_0(kr) - 4 J_2(kr) + J_4(kr) \right]
\nonumber \\
& + &
(9 \cos^2 \theta_t \sin^2 \theta_t  
+ \cos^2 \theta_t \sin^2 \theta_{t'} \nonumber\\
& & \,\, + 4 \cos \theta_t \sin \theta_t \cos \theta_{t'} \sin \theta_{t'} 
\nonumber \\
& & \,\, + \sin^2 \theta_t \cos^2 \theta_{t'}
+ 9 \cos^2 \theta_{t'} \sin^2 \theta_{t'})  \nonumber\\
& & \,\, \,\times \left[ J_0(kr) - J_4(kr) \right]
\nonumber \\
& + &
\left( \frac{3}{2}  \sin^4 \theta_t 
+ \sin^2 \theta_t \sin^2 \theta_{t'} 
+ \frac{3}{2} \sin^4 \theta_{t'} \right) \nonumber\\
&&\times\, \left[3 J_0(kr) + 4 J_2(kr) + J_4(kr) \right]
\Bigg\}.
\label{eq:offd}
\end{eqnarray}
We note that for $kr \gg 1$
\begin{equation}
\label{large bessel kr}
J_l(x) \longrightarrow \frac{1}{x} \sin \left( x - \frac{l \pi}{2} \right),
\end{equation}
and thus Eq.\ \ref{eq:offd} becomes
\begin{eqnarray}
\label{eq:offd:largekr}
R_{tt'} & = &  
1 - \frac{\alpha}{4} 
\left[2 + \cos(2 \theta_t)+ \cos (2 \theta_{t'}) \right] \nonumber \\
& & + \frac{\alpha^2}{8} 
\left(3 \cos^4 \theta_t + 
       2\cos^2 \theta_t \cos^2 \theta_{t'} + 3 \cos^4 \theta_{t'} \right).
\end{eqnarray}

We examine Eqs.\ \ref{eq:offd} and \ref{eq:offd:largekr} for three
different configurations of asymmetric beams that are the interesting
cases in the range of possible topologies.  Fig.\ \ref{figure
topologies} shows the three configurations.  In Case 1 the beams are
separated along the direction of the scan.  In Case 2 the beams are
separated perpendicular to the scan, and in Case 3 the separation
between the beams is parallel to the scan direction for one beam, and
perpendicular for the other beam.  For each one of these
configurations we fix the distance scale $r$ at two values, and plot
in Figure \ref{fig:fixr} 
the behavior of $R_{tt'}$ as a function of $k$.  We let
$k$ vary from $k=0$ up to $k_{\rm max} = 1/(2 v \tau)$ so that the
approximation $\alpha < 1$ is valid throughout.

\begin{figure}
\centerline{\psfig{file=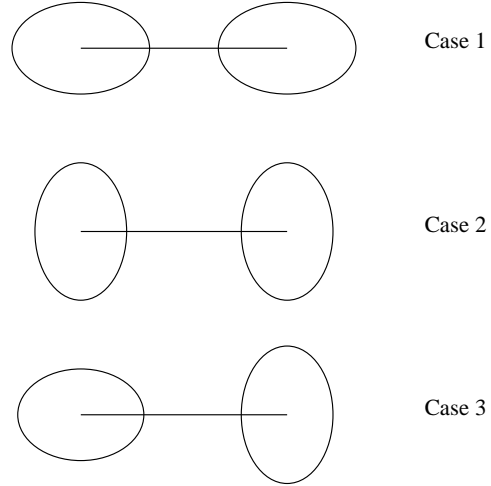,height=2.5in, angle=-90}}
\caption{Three beam configurations that contribute to the off-diagonal elements
  of the window function matrix. The configurations shown span the range
  of interesting topologies.}
\label{figure topologies}
\end{figure}

\begin{figure}
\centerline{\psfig{file=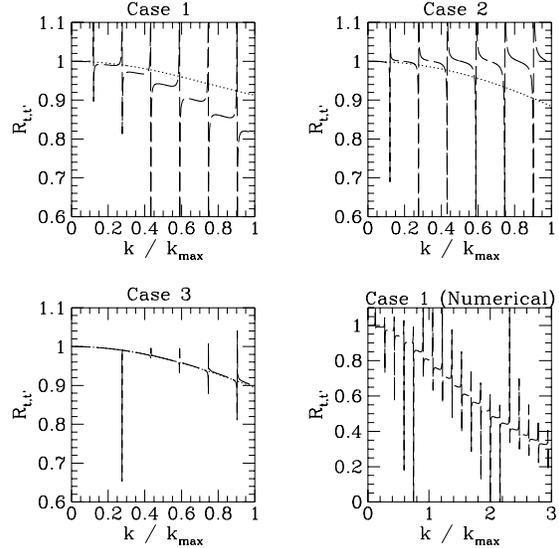,width=3.0in}}
\caption{$R_{tt'}(k)=W_{tt'}/B^2(k) J_0(kr)$ for the three different topologies
  illustrated in Fig.\ \ref{figure topologies}.  The upper left panel
  is Case 1, the upper right panel is Case 2, and the lower left panel
  is Case 3. In each panel $r = 1/k_{\rm max}$ (dotted) and $r= 20 /
  k_{\rm max}$ (long dashed), where $k_{\rm max} = 1/(2 v \tau )$. The
  periodic spikes are due to the shift between the zeros of Eq.\
  \ref{eq:offd} and $J_0(kr)$, and they can be ignored. Note that in
  these plates the vertical axis starts at $R_{tt'}=0.6$. In the lower
  right panel we show Case 1 again for a larger span in $k$ and for
  $r= 20 / k_{\rm max}$.  These results were obtained by numerically
  integrating Eq.\ \ref{eq:WtautoWzero}. In the range of overlap in
  $k$ the top left and bottom right panels agree.}
\label{fig:fixr}
\end{figure}

Case 1, shown in the upper left panel of Figure \ref{fig:fixr}, is
obtained by taking $\theta_t = \theta_{t'} = 0$ in Eq.\
\ref{eq:offd}.   
The periodic spikes in the value of $R_{tt'}$ are due to the slight
shift between the zeros of Eq.\ \ref{eq:offd} and $J_0(kr)$. Ignoring
these spikes we observe the following properties of $R_{tt'}$.  
Near $k = 0$, i.e., for very long wavelengths, there is no difference
whether the beams are elongated or not. As $k$ increases $R_{tt'}$ 
decreases for both values of $r$ indicating a smaller off-diagonal 
elements compared to the symmetric beam case. There is less
attenuation for small $r$ as compared to large $r$. 

To understand this we note that only those ${\bf k}$ vectors that are
directed near the line between the two beams contribute to the angular
average performed in Eq.\ \ref{eq:angular average}.  Since the
asymmetric beams are most widened in that direction large $k$'s are
effectively smeared (hence $R_{tt'}<1$). Larger attenuation is
obtained at large beam separation because a narrower range of wave
vectors, and only those that are aligned in the beam elongation
direction, contribute to the angular average. Within this range of
$k$'s the largest attenuation compared to the symmetric beam case is
$\simlt 20\%$. This limiting value is also obtained from Eq.\
\ref{eq:offd:largekr} with $\theta_t = \theta_{t'} = 0$ which gives
$R_{tt'}\approx 1 - \alpha + \alpha^2 = 0.8$.

Case 2 is obtained by taking $\theta_t = \theta_{t'} = \pi/2$ in Eq.\
\ref{eq:offd}.  The upper right panel of Fig.\ \ref{fig:fixr} shows
$R_{tt'}(k)$ for this case. Similar to Case 1, when $k =0$ we 
obtain the symmetric beam case, $R_{tt'}=1$. For $r=20/k_{\rm max}$
there is no change in $R$ as $k$ increases, and there is 
$\sim 10\%$ change for the $r=1/k_{\rm max}$ case. 

The physical explanation is similar to the reasoning in Case 1.  The
beams are narrowest in the direction of their separation. For large
$r$ only those $k$ vectors that are aligned with the narrow dimension
of the beam contribute to the angular average in Eq.\
\ref{eq:offd}. Hence there is effectively no difference between the
symmetric beam case and the asymmetric case. The large $r$ limit as
obtained from Eq.\ \ref{eq:offd:largekr} is $R_{tt'}=1$.  At small
$r$ values ${\bf k}$ vectors which are in the direction of the beam
elongation contribute to the angular average. In this case $R_{tt'}$
is smaller compared to either the symmetric case or to the large
$r$ value. 

Case 3 is obtained by taking $\theta_t = 0$ and $\theta_{t'} = \pi/2$
in Eq.\ \ref{eq:offd}.  The lower left panel of  Fig.\ \ref{fig:fixr}
shows $R_{tt'}(k)$ for this case. For arguments similar to the ones in
the previous cases we expect, and indeed observe, little difference
between the two values of $r$. We also observe an overall attenuation
at $k_{\rm max}$ of $\sim 10\%$. This agrees with the asymptotic value for
large $r$, $R_{tt'}(k)\approx 1 - \alpha/2 + 3/8\alpha^2 = 0.9$.

Within the approximation $kv\tau = \ell/\ell_{\tau} \leq 0.5$ the
largest reduction in the off-diagonal elements occurred
for case 1, and for large values of $r$. For this case we continue the
analysis for larger values of $kv\tau$ by numerically integrating Eq.\
\ref{eq:WtautoWzero}. The result of this numerical integration is
shown in the lower right panel of Fig.\ \ref{fig:fixr}. We find that the
analytic approximation is excellent for $kv\tau = \ell/\ell_{\tau} \leq
0.5$ and that when $\ell/\ell_{\tau} = 1$ the magnitude of these particular
off-diagonal elements is about 50\% of their values in the symmetric
beam case. 

We emphasize that we are considering here the effect of the asymmetric
beam {\em relative} to the symmetric beam case.  In Case 1 the
asymmetric beam causes the greatest relative effect at large separation.
However, the {\em absolute} contribution of these matrix elements to the
signal at large $\ell$ will be small because large separations
sample mainly small $\ell$ values (that is, the matrix
elements are suppressed for $kr\simeq\ell r\gg1$). We thus find
that for the $\ell$ range where beam elongation is important, i.e.,
large $\ell$, the matrix elements are not attenuated significantly
compared to the symmetric beam case, certainly less than the 50\%
reduction observed for the most severe case at $\ell/\ell_\tau=1$. Based
on the arguments of Section \ref{section off diagonal elements} we
expect a similar increase in the relative errors $\delta C_{\ell}/
C_{\ell}$.  The change in the relative error expected for any one of the
cosmological parameters will be similar or smaller, and thus the major
conclusions of Table \ref{table cosmological parameters} will not
change.

\section{Discussion}
\label{section discussion}

Our point source response analysis shows that using a combination of
bolometer time constant, scan speed, and beam size such that
$L=\sigma/v\tau = \ell_{\tau}/\ell_{\sigma} \simgt 2.5 $ results in
relatively small increase in the effective beam size, less than 16\%,
and the spatial phase shift is less than 0.2 $\sigma$.  Because the
window function encodes information from the nominal beam dimension as
well as from the enlarged one the effect of a smaller $L$ is less
pronounced on the window function than it is on the point source
response in the scan direction.  In the Gaussian approximation the
fractional effective increase in beam size derived from the window
function analysis is a factor of two smaller than that derived from
the point source response (Eq.\ \ref{new 2d width}).

Our analysis of the performance of Planck-HFI indicates that choosing
$L \sim 0.5$ will not be detrimental to the ability to extract
cosmological parameters.  This is a consequence of HFI's small beam
size and high sensitivity. The HFI, with its nominal $\ell_{\sigma}
=1/\sigma$ which range from 725 to 1800, will determine the CMB power
spectrum with high accuracy over a very broad range of $\ell$.  A
degradation of its performance at high $\ell$ values does not diminish
its capability significantly. It will generally be the case that the
ability of any experiment with $\ell_{\sigma} \simgt 800$ to extract
cosmological parameters will not be affected significantly with $L
\simgt 0.5$, if the universe is any variant of the CDM model.

There are strong arguments for maintaining high resolution capability
for a CMB experiment. We do {\em not} know that the universe is any
variant of the cold dark matter model, or indeed that the fluctuations
are Gaussian. High resolution will undoubtedly be important to test
this latter assumption.  Also, the ability to remove foregrounds and
to detect point sources relies to some extent on high $\ell$
information and the ability to observe small-scale non-Gaussian
structures.

If we are interested in achieving the maximum possible resolution, we
wish to minimize a quantity like the effective 2-d beamwidth
$\sigma_{2d}^2\simeq\sigma^2(1+L^{-2}/2)$.  On the other hand, other
considerations will make a {\em smaller} $L$ more desirable. A fast
scan speed increases the scale over which instrument noise is
uncorrelated. Such broad temporal and/or angular scales can be used to
better characterize the instrument's noise, to increase the $\ell$
range to which the experiment is sensitive, or for a better recovery
of the underlying two dimensional sky signal \cite{tegmark1}.  The
form of $\sigma_{2d}$ argues that we quickly reach a regime of
diminishing returns as $L\gg 1$. For a given experiment the final
determination of $L$ depends on a global optimization which includes
the expected instrument noise characteristics and the targeted
science.  Our analysis indicates that $1 \simlt L\simlt 2.5$ provides
a range over which the small angular scale performance is not affected
significantly, while allowing flexibility in tuning either the
bolometer response time and/or the scan speed.  For the HFI in
particular, relaxing the value by a factor of $\sim 2$, to $L \sim
1.5$, may increase the dynamic range over which the instrument and
satellite parameters can be varied.  Some experiments may choose to
use even smaller values of $L$.

\section*{Acknowledgments}

We acknowledge and thank Viktor Hristov and Phil Mauskopf who first
investigated the relation between the scan speed, beam size and detector
response time. Phil Mauskopf has also independently suggested the use of
the temporal window function.  We thank Pedro Ferreira for discussions,
and Paul Richards for comments on the manuscript. We are grateful to
Dick Bond for the use of the code that calculates the errors on the
determination of the cosmological parameters. S.\ H.\ and A.\ H.\ J.\ 
acknowledge support by NASA grant NAG5-3941 and by NSF cooperative
agreement AST-9120005.  S.\ H.\ was also supported by NASA grant
NAG5-4454 and A.\ H.\ J.\ by NAG5-6552. E.\ S.\ is supported by an NSF
graduate student fellowship.

\end{document}